\documentclass[preprint,prb,epsf,graphics,psfig,superscriptaddress]{revtex4}

\usepackage{graphicx,graphics} % Include figure files
\usepackage[dvips,bookmarks=false]{hyperref}
\usepackage{graphicx}
\usepackage{dcolumn}
\usepackage{eucal}
\usepackage[dvips]{epsfig}

\begin{document}
\title{Optimisation of quantum Monte Carlo wave function: steepest descent method}
\author{M. Ebrahim Foulaadvand }
\affiliation{ Department of Physics, Zanjan University, P.O. Box
45196-313, Zanjan, Iran}
\affiliation{Computational Physical
Sciences Laboratory, Department of Nano-Science, Institute for
Research in Fundamental Sciences (IPM), P.O. Box 19395-5531, Tehran,
Iran}
\author{Mohammad Zarenia}
\affiliation{Department of Physics, Tarbiat Modares University, P.O.
Box 14115-111, Tehran, Iran }
\date{\today}

\begin{abstract}

We have employed the steepest descent method to optimise the
variational ground state quantum Monte Carlo wave function for He,
Li, Be, B and C atoms. We have used both the direct energy
minimisation and the variance minimisation approaches. Our
calculations show that in spite of receiving insufficient attention,
the steepest descent method can successfully minimise the wave
function. All the derivatives of the trial wave function respect to
spatial coordinates and variational parameters have been computed
analytically. Our ground state energies are in a very good agreement
with those obtained with diffusion quantum Monte Carlo method (DMC)
and the exact results.

\end{abstract}
\maketitle
%{\bf PACS.}

\section{Introduction}

Quantum Monte Carlo (QMC) method has constituted an
efficient and powerful numerical method for solving time-independent
many-body Schr\"{o}dinger equation mainly in chemistry and solid
state physics
\cite{hammond,ceperley96,umrigar99,foulkes,mcmillan,ceperley77,fahy,raghavachari,needs,neekamal}.
Among various approaches to QMC namely, random walk, diffusion,
Green-function etc, variational quantum Monte Carlo (VMC) has been
extensively studied in recent years \cite{foulkes}. In VMC method, a
parameterized many-body trial wave function is optimised according
to Raleigh-Ritz variation principle. In practice, this task is done
utilizing a numerical algorithm for optimisation the parameters.
Various algorithms have been proposed and implemented in the
framework of QMC such as Newton \cite{lin,umrigar05,sorella05}, steepest descent (SD)
\cite{huang90,huang96,huang99}, perturbative optimisation
\cite{scemma,toulouse} and linear optimization method \cite{toulouse,umrigar07,toulouse08}. The wave-function optimisation is
implemented via two schemes namely {\it energy minimisation} and
{\it variance minimisation}. These methods have their own merits and
disadvantages. A basic task in VMC is the evaluation of first and
second derivatives of the local energy $E_L=\frac{H \Psi}{\Psi}$
respect to variational parameters and spatial coordinates or a
combination of them ($\Psi$ is the trial wave function). Despite normally
the first derivative is analytically evaluated and the second
derivatives are calculated numerically \cite{luchow} there are papers in which second
derivatives are also calculated analytically \cite{umrigar05}. Numerical evaluation of second derivatives
causes a systematic error into the problem. To the best of our knowledge, the
SD method has only been utilized in the variance minimisation
approach \cite{huang96}. Our objective in this paper is to show that
implementation of the SD method in the direct approach of energy
minmisation yields reasonable results. We report our results for the
ground state energies of He, Li, Be, B and C atoms and compare them
to the results in the literature obtained by other methods.

\section{ Variational wave function and steepest descent optimisation method  }

\subsection{Theoretical background}

Let us briefly explain the basic ingredients of the VMC method. In
the VMC method a trial many-body wave function
$\Psi(\vec{R},\{c_m\})$ containing a set of $M$ variational
parameters $c_1,c_2,\cdots,c_M$ is considered. $\vec{R}$ denotes the
position set of electrons. We confine ourselves to Born-Openheimer
approximation in which the nuclei are assumed static and only the
electronic degrees of freedom are taken into account. The parameters
${c_m}$ are varied according to the Raleigh-Ritz variation procedure
so as to minimize the variational energy $E(\{c_m\})$ defined as
follows:

\begin{equation}
E(\{c_m\})=\frac{\int\Psi^*(\vec{R},\{c_m\})H\Psi(\vec{R},\{c_m\})d\vec{R}}{\int\Psi^*(\vec{R},\{c_m\})\Psi(\vec{R},\{c_m\})d\vec{R}}
\end{equation}

Where $H$ is the many body system Hamiltonian. We ignore
relativistic correction and take the Hamiltonian as follows (in Hartree atomic units):

\begin{equation}
H=-\frac{1}{2}\sum_i \nabla_i^2 -\sum_{i,I} \frac{Z_I}{r_{iI}} +
\sum_{i<j}\frac{1}{r_{ij}}
\end{equation}

Small letters refer to electrons and capital ones to nuclei. $Z_I$
is the electric charge of the $I$-th nucleus and $r_{ij}$ denotes
the distance between electron $i$ and electron $j$ whereas $r_{iI}$
denotes the distance between electron $i$ and nucleus $I$. Moreover,
we restrict ourselves to real-valued wave function $\Psi$ and omit
the complex conjugate symbol afterwards. By introducing a local
energy $E_L=\frac{H\Psi}{\Psi}$ and a normalized probability
distribution function
$p(\vec{R},\{c_m\})=\frac{\Psi^2(\vec{R},\{c_m\})}{\int\Psi^2(\vec{R},\{c_m\})d\vec{R}}$
We recast equation (1) in the following form:

\begin{equation}
E(\{c_m\})=\int p(\vec{R},\{c_m\})E_L d \vec{R}
\end{equation}

It is now possible to approximate the integral by the standard Monte
Carlo procedure:

\begin{equation}
E(\{c_m\})=lim_{N_{MC}\rightarrow \infty}
\frac{1}{N_{MC}}\sum_{s=1}^{N_{MC}} E_{L,s}
\end{equation}

$E_{L,s}$ denotes the local energy of the $s$-th sample of the
configuration-space and $N_{MC}$ is the number of Monte Carlo
sampling for evaluation of the integral.

\subsection{Optimisation of the wave function: steepest descent method}

The next step is to find the optimal values of the parameters which
minimise the objective function i.e.; the variational energy $E$
\cite{sorella,schautz,toulouse}. There exists numerous optimisation
methods such as Newton \cite{lin,umrigar05,myung}, steepest descent
\cite{huang90,huang96,huang99}, conjugate gradient etc in the
literature. Here we focus on the simplest of them and show that
despite simplicity algorithm is capable of exhibiting a satisfactory
performance despite not receiving much attention. We briefly recall
the main ingredient of this method. Having numerically computed the
energy $E$ for a given set of parameters in (4), we iteratively
update the values of the parameters according to the following
procedure:

\begin{equation}
{\bf c} ^{k+1}={\bf c} ^{k} - a {\bf g}^k
\end{equation}

The vector ${\bf c}^\dag=(c_1,c_2,\cdots,c_M)$ denotes the
parameters, $k$ is the iteration step and $a$ denotes the constant
of the SD method. The vector ${\bf g}^\dag=(\frac{\partial
E}{\partial c_1},\frac{\partial E}{\partial
c_2},\cdots,\frac{\partial E}{\partial c_M})$ is the gradient vector
of energy respect to the parameters. We note that in some cases we
should vary the SD constant in each iteration step to get the
desired optimum value. In order to utilize SD method, we should
evaluate the energy gradient vector. This has been done in details
in \cite{lin}. We only quote the result:

\begin{equation}
\frac{\partial E}{\partial c_m}=lim_{N_{MC}\rightarrow \infty}
\frac{2}{N_{MC}}\sum_{s=1}^{N_{MC}} \{ (\frac{\partial ln
\Psi}{\partial c_m})_s (E_{L,s} -E ) \}
\end{equation}

In eq. (6) $(\frac{\partial ln \Psi}{\partial c_m})_s$ denotes the
logarithmic derivative of wave function evaluated in the $s$-th MC
configuration. If the constant $a$ is appropriately chosen the
sequence ${\bf c} ^{k}$ converges to ${\bf c} ^{*}$ after some
iterations.

\subsection{Trial wave function and its parameters}

We wish now to introduce the structure of the trial ground state
wave function we have implemented in our calculations for simple
atoms. We have taken the following well-known form for
$\Psi$\cite{boys,schmidt}:

\begin{equation}
\Psi=D^\uparrow D^\downarrow e^J
\end{equation}

In which $D^\uparrow$ and $D^\downarrow$ are up-spin and down-spin
Slater determinants and $J$ is the Jastrow factor. The number of
spatial orbitals $N_{up}(N_{down})$ in the construction of Slater
determinant $D^\uparrow (D^\downarrow)$ equals the number of spin up
(down) electrons and depends on the atom we consider. Note that
$N_{up} + N_{down}=N$ where $N$ is the number of electrons in the
atom. For the basis set in the construction of up and down Slater
determinants we have used a variant of Slater-type $s$ and $p$
orbital as follows \cite{garciaI,garciaII}:

\begin{equation}
\phi_s({\bf r})= \sum_{k=1}^PC_ke^{-\zeta_kr} + \sum_{k=1}^PC'_k r
e^{-\zeta'_kr}
\end{equation}

\begin{equation}
\phi_{p_x}({\bf r})= \sum_{k=1}^PxD_k e^{-\xi_kr}
\end{equation}

Analogous definitions goes for $p_y$ and $p_z$ orbitals. We have set
$P=3$ in all our calculations. Henceforth, the parameters are
$C_k,C'_k,D_k,\zeta_k,\zeta'_k$ and $\xi_k$ where $k=1,2,3$. For the
Jastrow factor we have taken the following form:

\begin{equation}
J= \sum_{i<j}U_{ij}
\end{equation}

The sum goes over all the particles (electrons) and $U_{ij}$ has the
following dependence on distances:

\begin{equation}
 U_{ij}=\sum_{mno}C_{mno}[(\frac{r_i}{1+r_i})^m(\frac{r_j}{1+r_j})^n +
(\frac{r_i}{1+r_i})^n(\frac{r_j}{1+r_j})^m](\frac{r_{ij}}{1+r_{ij}})^o
\end{equation}

$r_i$ is the distance between electron $i$ and the nucleus, $r_{ij}$
is the distance between electrons $i$ and $j$. Exponents $m,n,o$ are
positive integers and the sum over $mno$ denotes the sum over given
values of these integers. We adopt the following choice of integers
\cite{schmidt} $m,n,o$:
$$\{(0,0,1),(0,0,2),(0,0,3),(0,0,4),(2,0,0),(3,0,0),$$
\begin{equation}
(4,0,0),(2,2,2),(2,0,2)\}
\end{equation}

Equation (11) includes electron-electron correlations (terms with $m=n=0$), electron-nucleus correlations
($o=0$ as well as one of $m$ or $n$ zero) and also electron-electron-nucleus correlations ((2,0,2) and (2,2,2) terms).
Here we have considered the simplest choices compatible with electron-electron and electron-nucleus cusp conditions.
The origin of three body correlation terms in (11) stems in the back flow correlation firstly suggested by Feynman and
Cohen \cite{feynman}. We refer the readers for more details to reference [25].
The Jastrow function has nine independent parameters $C_{mno}$. Each
$s$ type orbital contains twelve parameters and in a $p$ orbital we
have six parameters. We note that after imposing electron-nucleus
cusp conditions, one parameter from each $s$ orbital will be fixed.

\subsection{Variance minimisation method}

In the preceding sections, we outlined the basics of energy
minimisation method. In this method, we minimise the variational
energy. In recent years, an alternative scheme the so-called {\it
variance minimisation} has been introduced
\cite{umrigar88,umrigar05} and has become one of the most frequently
used method in the literature. This method has shown to provide some
advantages over the straightforward energy minimisation. We now
briefly review this method. Instead of energy, we minimise the
variance of the local energy $E_L$ \cite{coldwell,bartlett}:

\begin{equation}
\sigma^2=\frac{\int \Psi^2(\vec{R},\{c_m\})(E_L-E)^2 d \vec{R}}{\int
\Psi^2(\vec{R},\{c_m\})d \vec{R}}= \langle (E_L -E)^2\rangle
\end{equation}

All the other steps are analogous to those in the energy
minimisation method. To implement the SD procedure we only should
replace the energy gradient vector with the variance gradient
vector. Derivatives of $\sigma^2$ respect to parameters have been
evaluated in \cite{umrigar05}. Here for simplicity we use the following expression which ignores the change of the wave function \cite{umrigar05}:

\begin{equation}
\frac{\partial\sigma^2}{\partial c_m}=2\langle \frac{\partial
E_L}{\partial c_m}(E_L-E)\rangle
\end{equation}

More concisely the above approximation corresponds to underweighted variance minimisation method. The average is taken with the
normalised probability function $p(\vec{R},\{c_m\})=\frac{\Psi^2(\vec{R},\{c_m\})}{\int\Psi^2(\vec{R},\{c_m\})d\vec{R}}$.
We can approximate the average in (14) by a sum in MC approach. Note
that when implementing this method, we have to replace $E$ with
$\sigma^2$ in the gradient vector ${\bf g}^\dag$ in equation (5). In
the next section our results will be reported. All the computational
details of the calculations are explained in the appendix.

\section{Application to atoms and discussion}

We have implemented the steepest descent optimising method to find
the ground state energy and wave function of atoms He, Li, Be, B and
C by two approaches of energy and variance minimisation. Let us now explain our procedure of energy minimisation. It consists of three steps: {\it anticipating} the variational parameters, {\it finding the optimised value} of steepest descent parameter $a$ and eventually the {\it fine tuning} of variational energy. Step one begins with random initialization of the parameters values. Initial values of Jastrow parameters are randomly chosen in vicinity of zero. We then set the SD constant $a$ to a rather high value say $a=0.1$. Next we proceed with some iterations of (5) until the variational energy reaches approximately to the exact ground state energy. This comprises step one. During this step integrals in (1) are evaluated by the standard MC Metropolis method. Each MC move consists of a random selection of an electron and displace it from its position by the vector $\vec{\delta}$. The move size which is the length of $\vec{\delta}$ is randomly chosen (uniformly) from the interval $[0,\delta_{max}]$. The direction of $\vec{\delta}$ is uniformly chosen between zero and $2\pi$. We took the number of Monte Carlo steps $N_{MC}$ equal to $3\times 10^{5}$. We discard the first $10000$ steps to ensure reaching equilibrium. Averages are separated by $20$ MC steps to suppress the effects of correlations among generated MC configurations. The MC maximum move size $\delta_{max}$ has been typically $0.3$ (Hartree atomic units) with the acceptance ratio around $70$ percent. At the end of step one, which normally takes $5-6$ iterations, variational parameters should have reached to the vicinity of their ultimate values. We put their latest values in the code and re run it. This is the beginning of step two. We then proceed with some iterations until the iteration series of the variational energy begins to diverge. This shows that by the current value of the SD parameter we can no more reach the true energy. Here we reduce $a$ to a smaller value say one order of magnitude smaller and repeat the procedure until the iteration series of energy begins to diverge or strongly oscillates.
We repeat this $a$ reduction procedure until further reduction of the SD constant $a$ does not lead to divergence of energy iteration series. Normally after $4-5$ repetitions we achieve our aim and $a$ reaches to a value of the order $10^{-5}$. This marks the end of step two and by now we have an iteration energy series. In figure (1) we have depicted such series of Be ground state energy obtained in the method explained above. Corresponding series for other atoms are similar in nature.

\begin{figure}
\centering
\includegraphics[width=7.5cm]{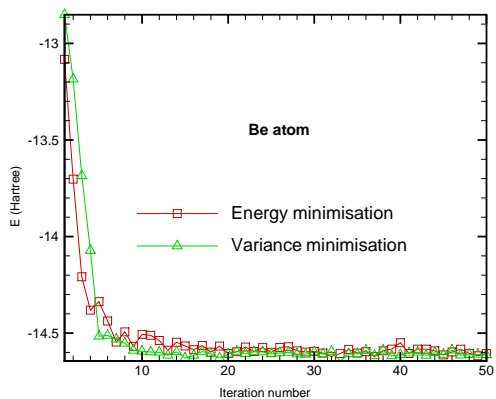}
\caption{ Ground state energy of Be in two methods of energy and
variance minimisation. } \label{fig:bz2}
\end{figure}

It is seen that after roughly $20$ iterations we reach a steady state
regime. Next in figure (2) we exhibit the dependence of absolute
value of $\nabla_c E$ on the iteration number.

\begin{figure}
\centering
\includegraphics[width=7.5cm]{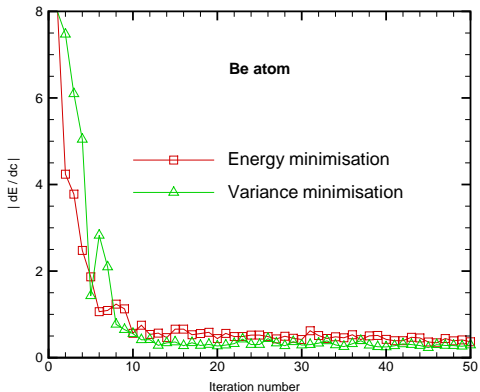}
\caption{   $|\nabla_c E|$ vs iteration number.} \label{fig:bz2}
\end{figure}

We see that $|\nabla_{{\bf c}} E|$ tends to a small value for a sufficient number of iterations. Theoretically it should goes to zero
but due to numeric computations it does not approach to zero. Finally in third step, in which the fine tuning of energy in performed, we choose the variational parameters obtained from that iteration in the series which has the smallest energy and re run the code for these values of parameters for a longer MC run of $N_{MC}=2\times 10^{6}$ steps to find the energy in a fine tune manner. All the reported values in table I have been obtained in this way. In figure (3) the energy error is shown for both methods of energy and variance optimisation. We recall that the error has been obtained from the following formula:

\begin{equation}
error= \sqrt{\frac{\frac{1}{N_{MC}}\sum_{i=1}^{N_{MC}}
[E_L(\vec{R_i})]^2-E^2}{N_{MC}-1}}
\end{equation}

\begin{figure}
\centering
\includegraphics[width=7.5cm]{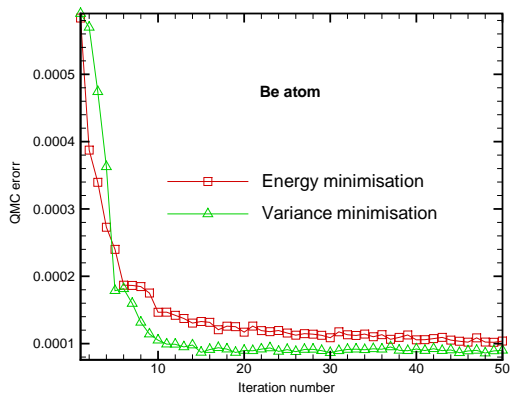}
\caption{ QMC error vs iteration number.} \label{fig:bz2}
\end{figure}

We see that the error is lower in variance minimisation method. In
both methods, the error decreases as we increase the iteration
number. In table I, we report the ground state energies we have
obtained and compare them both to the existing computational results
in the literature obtained by other methods
\cite{lin,schmidt,casula} and the exact ones \cite{chakravorty}. The
comparison shows that the steepest descent method is capable of
minimising the energy to a very good precision. In fact our results
by energy minimisation method is in most of the cases even better
than those reported in \cite{schmidt,lin}. Besides Be, in all the
atoms, our energy is comparable to the DMC energy. This marks the
efficiency of the steepest descent method at least for light atoms.
We should like to emphasize that simplicity is the main merit of our
approach which can turn it into an efficient method al least for simple atoms.
Our results obtained by variance minimisation method is less
favourable in comparison to our energy minimisation ones. however,
the error in the energy minimisation method is larger. Comparison of
our variance minimisation results to those in \cite{lin} (which have
been obtained by VMC in Newton optimisation method ) shows that the
results of \cite{lin} is slightly better than ours. We have also
compared our results to the recent paper of Brown et al \cite{brown}
in which besides single determinant, multideterminant trial wave
function have been employed. It is seen that the accuracy of our
results is comparable to those exhibited in \cite{brown}.

\section{Summary and Concluding Remarks}

In summary we have applied the steepest descent optimisation method
to optimise the parameters of the QMC many-body wave-function in
some light atoms. Two schemes of energy and variance minimisation
have been implemented. The key features to achieve the correct
minimum is to vary the SD constant $a$ appropriately. Our results
are in a well agreement with exact results and those obtained by
DMC. We note that all the derivatives of the trial wave function
respect to spatial coordinates and variational parameters have been
analytically calculated.
\begin{table}
  \centering
  \caption{Variational energy with error bar for atoms He to C (All the energies are in Hartree).\\}\label{table2}

\begin{tabular}{c c c c c c}
\hline \hline
   % after \\: \hline or \cline{col1-col2} \cline{col3-col4} ...
     &   He   &   Li   &   Be   &  B   &   C  \\
  \hline
  $E$(energy minimisation) & -2.9037(4)&-7.4780(4)&-14.648(1)&-24.640(9)&-37.831(8)\\
  $E$(variance minimisation)& -2.9031(2)&-7.4757(3)&-14.6443(9)&-24.6244(3)&-37.807(5)\\
  $E_0$(Ref\cite{chakravorty}) &-2.903719&-7.47806&-14.66736&-24.65391&-37.8450\\
  $E_{VMC}$(Ref\cite{lin}) & -2.903717(8)&-7.47722(4)&-14.6475(1)&-24.6257(1) &-37.8116(2) \\
  $E_{VMC}$(Ref\cite{schmidt}) & -2.9029(1) &-7.4731(6)& -14.6332(8)&-24.6113(8)&-37.7956(7)\\
  $E_{DMC}$(Ref\cite{casula}) &-2.903719&-7.4780(2)&-14.6565(4)&-24.63855(5)&-37.8296(8)\\
  $E_{VMC}$(Ref\cite{brown}) &no report&-7.47683(3)&-14.6311(1)&-24.6056(2)&-37.8147(1)\\

  \hline
  \hline
\end{tabular}

\end{table}

\section{acknowledgement}

We highly appreciate very useful discussions with Mehdi Neek Amal.
MEF is thankful to N. Arshado Do'leh for his useful helps.

\section{Appendix}

In this section we give some details of the manipulations for the
evaluation of the integrals (3) and (14). In evaluation of these
integrals, we have analytically calculated two basic quantities
$E_L,\frac{\partial ln \Psi}{\partial c_m}$ and $\frac{\partial
E_L}{\partial c_m}$. Let us first consider $E_L$. According to its
definition we have:

\begin{equation}
E_L=\frac{H
\Psi}{\Psi}=-\frac{1}{2}\sum_i\frac{\nabla_i^2\psi}{\psi}+V
\end{equation}

The first term is kinetic energy and $V$ represents the potential
energy. Calculating $V$ is straightforward. To calculate the kinetic
energy $KE$ we rewrite it in the following form \cite{kent}:
\begin{equation}
KE=\sum_i-\frac{1}{2}[{\nabla_i^2\ln\Psi+(\overrightarrow{\nabla}_i
\ln\Psi)^2}]
\end{equation}

Concerning the form of the trail wave function, $\Psi=D^\uparrow
D^\downarrow e^J$ we have:

\begin{equation}
 \overrightarrow{\nabla}_i\ln\Psi=\frac{1}{D^\uparrow
}\overrightarrow{\nabla}_i D^\uparrow + \frac{1}{D^\downarrow
}\overrightarrow{\nabla}_i D^\downarrow +\overrightarrow{\nabla}_i J
\end{equation}
and

$$\nabla_i^2\ln\Psi=-(\frac{1}{D^\uparrow }\overrightarrow{\nabla}_i
D^\uparrow)^2 + \frac{1}{D^\uparrow }\nabla_i^2
D^\uparrow-(\frac{1}{D^\downarrow }\overrightarrow{\nabla}_i
D^\downarrow)^2 +$$

\begin{equation}
\frac{1}{D^\downarrow }\nabla_i^2 D^\downarrow +\nabla_i^2 J
\end{equation}

We note that the entry $D_{ij}$ of any determinant $D$ equals
$\phi_i({\bf r}_j)$  in which the orbital $\phi_i$ is the $i$th
orbital in the construction of D. To evaluate the matrix
$\overrightarrow{\nabla}_iD$ we only have to replace the $i$th
column of matrix D $C_i$ by a new column
$\tilde{C}_i=(\overrightarrow{\nabla}_i \phi_1({\bf
r}_i),\overrightarrow{\nabla}_i \phi_2({\bf r}_i),\cdots,
\overrightarrow{\nabla}_i \phi_{N_D}({\bf r}_{N_D}))^{\dagger}$. The
matrix $\nabla^2_i D$ is analogously constructed but with
$\tilde{C}_i=(\nabla^2_i \phi_1({\bf r}_i),\nabla^2_i \phi_2({\bf r}_i),\cdots,\nabla^2_i \phi_{N_D}({\bf r}_{N_D}))^{\dagger}$. $N_D$ is the dimension of D.\\

The next quantity to evaluate is $\frac{\partial ln \Psi}{\partial
c_m}$. Some straightforward calculations yields:

\begin{equation}
\frac{\partial\ln\Psi}{\partial c_m}=\frac{1}{D^\uparrow
}\frac{\partial D^\uparrow }{\partial c_m }+\frac{1}{D^\downarrow
}\frac{\partial D^\downarrow }{\partial c_m }+\frac{\partial
J}{\partial c_m}
\end{equation}

The derivative of a Slater determinant respect to $c_m$ equals a sum
of $N_D$ determinants. The $i$th term of this sum is the determinant
D with its $i$th column $C_i$ replaced with column
$\tilde{C}_i=(\frac{\partial \phi_1({\bf r}_i)}{\partial
c_m},\frac{\partial \phi_2({\bf r}_i)}{\partial c_m},
\cdots,\frac{\partial \phi_{N_D}({\bf r}_i)}{\partial
c_m})^{\dagger}$.

Eventually in order to evaluate $\frac{\partial E_L}{\partial c_m}$
we proceed as follows (starting with (16)):

\begin{equation}
\frac{\partial E_L}{\partial
c_m}=-\frac{1}{2}\sum_i\frac{\partial\nabla^2_i\ln\Psi}{\partial
c_m}-\sum_i\overrightarrow{\nabla}_i\ln\Psi\cdot\frac{\partial\overrightarrow{\nabla}_i\ln\Psi}{\partial
c_m}
\end{equation}

The terms containing derivatives respect to $c_m$ can be evaluated
as follows:

$$\frac{\partial\overrightarrow{\nabla}_i\ln\Psi}{\partial
c_m}=-\frac{1}{(D^\uparrow)^2}\frac{\partial D^\uparrow}{\partial
c_m}\overrightarrow{\nabla}_i D^\uparrow
+\frac{1}{D^\uparrow}\frac{\partial \overrightarrow{\nabla}_i
D^\uparrow}{\partial c_m} -$$

\begin{equation}
\frac{1}{(D^\downarrow)^2}\frac{\partial D^\downarrow}{\partial
c_m}\overrightarrow{\nabla}_i D^\downarrow
+\frac{1}{D^\downarrow}\frac{\partial \overrightarrow{\nabla}_i
D^\downarrow}{\partial c_m} +\frac{\partial
\overrightarrow{\nabla}_i J}{\partial c_m}
\end{equation}

The second term gives the following expression:

$$
\frac{\partial\nabla^2_i\ln\Psi}{\partial
c_m}=\frac{2}{(D^\uparrow)^3}\frac{\partial D^\uparrow}{\partial
c_m}|\overrightarrow{\nabla}_i
D^\uparrow|^2-\frac{2}{(D^\uparrow)^2}\frac{\partial\overrightarrow{\nabla}_i
D^\uparrow}{\partial c_m}\cdot\overrightarrow{\nabla}_i D^\uparrow
$$

$$-\frac{1}{(D^\uparrow)^2}\frac{\partial D^\uparrow}{\partial
c_m}\cdot\nabla^2_i D^\uparrow+ \frac{1}{D^\uparrow}\frac{\partial
\nabla^2_i D^\uparrow}{\partial c_m}$$

$$+\frac{2}{(D^\downarrow)^3}\frac{\partial D^\downarrow}{\partial
c_m}|\overrightarrow{\nabla}_i D^\downarrow|^2-
\frac{2}{(D^\downarrow)^2}\frac{\partial\overrightarrow{\nabla}_i
D^\downarrow}{\partial c_m}\cdot\overrightarrow{\nabla}_i
D^\downarrow -$$

\begin{equation}
\frac{1}{(D^\downarrow)^2}\frac{\partial D^\downarrow}{\partial
c_m}\cdot\nabla^2_i D^\downarrow+
\frac{1}{D^\downarrow}\frac{\partial \nabla^2_i
D^\downarrow}{\partial c_m}+ \frac{\partial \nabla^2_i J}{\partial
c_m}
\end{equation}

Note that in equations (22) and (23) to evaluate $\frac{\partial
\nabla^2_i D}{\partial c_m}$ and $\frac{\partial
\overrightarrow{\nabla}_i D }{\partial c_m}$ , we should first
evaluate $\nabla^2_iD$ and $\overrightarrow{\nabla}_i D$ and then
implement the derivatives respect to $c_m$.

%\end{multicols}

\end{document}